\documentclass[a4paper,twocolumn,aps,prl,longbibliography]{revtex4-2}

\newcommand{\eq}[1]{\begin{equation}#1\end{equation}}
\newcommand{\dd}{\mathrm{d}}
\newcommand{\ee}{\mathrm{e}}

\newcommand{\elk}{\varepsilon_{\ell,k}}
\newcommand{\clk}{\chi_{\ell,k}}

\newcommand{\Tr}{\mathrm{Tr\,}}

\newcommand{\ehcft}{\mathcal{\hat H}_{\mathrm{\tiny{CFT}}}}
\newcommand{\ehlcft}{\mathcal{\hat H}_{\ell,\mathrm{\tiny{CFT}}}}

\newcommand{\twn}[1]{\langle \mathcal{T}_n(#1)\rangle}
\usepackage{amsmath}
\usepackage{amsfonts}
\usepackage{graphics}
\usepackage{graphicx}
\usepackage{psfrag}
\usepackage{color}
\usepackage{colordvi}
\usepackage{bbm}
\usepackage{braket}
\usepackage{physics}

\usepackage{hyperref}

\begin{document}

\title{Entanglement Hamiltonian of a nonrelativistic Fermi gas}

\author{Viktor Eisler}

\affiliation{
Institute of Theoretical and Computational Physics,
Graz University of Technology, Petersgasse 16, 8010 Graz, Austria}

\begin{abstract}
We study the entanglement Hamiltonian for a spherical domain in the ground state of a
nonrelativistic free-fermion gas in arbitrary dimensions. Decomposed into a set of radial
entanglement Hamiltonians, we show that the entanglement spectrum in each sector
is identical to that of a hopping chain in a linear potential, with the angular momentum
playing the role of the subsystem boundary. Furthermore, the eigenfunctions follow from
a commuting differential operator that has exactly the form predicted by conformal
field theory. Rescaled by the radial Fermi velocity, this operator gives a perfect approximation
of the entanglement Hamiltonian, except for large angular momenta that belong to the edge regime in the
analogous gradient chain. One thus finds that the conformal field theory result
becomes asymptotically exact only in one dimension.

\end{abstract}

\maketitle

Entanglement plays a key role in characterizing the distinct phases of quantum matter
in ground states of many-body systems \cite{AFOV08,CCD09,ECP10,Laflo16}.
The intricate nature of quantum correlations is encoded in the reduced density matrix of a subsystem,
or equivalently, written in an exponential form, in the entanglement Hamiltonian (EH) \cite{DEFV22}.
One of the most remarkable property that has been uncovered in a broad range of many-body systems
is the locality of the EH \cite{DEFV22}. Its precise structure is, however, not only of theoretical interest, but also
fundamental to novel techniques aiming at a more efficient spectroscopy and tomography
of the reduced density matrix in quantum simulators \cite{DVZ18,Kokailetal21a,Kokailetal21b,Zacheetal2022}.
These protocols perform a variational learning of the EH from the available measurement data,
and have recently led to breakthrough results in ion-trap \cite{Joshietal23} and cold-atom
\cite{Redonetal23} experiments.

In the above mentioned applications, it is crucial to have an educated ansatz for the EH,
which is mainly guided by the Bisognano-Wichmann theorem
of relativistic quantum field theory \cite{BW75,BW76}. This provides the EH of a half-infinite system via the physical
energy density, weighted by an inverse temperature that increases linearly from the entanglement cut,
and is valid in arbitrary dimensions. Generalizations to different geometries exist within conformal
field theory (CFT), and yield again a local result with a modified weight function \cite{HL82,CHM11,WKZV13,CT16}.

In practice, however, one typically faces a problem, where Lorentz invariance is explicitly broken
by the presence of a lattice. Although quantum field theory may still provide an effective low-energy
description, it is crucial to address the robustness of the results for the EH.
In particular, the analytical solution for a free-fermion chain shows \cite{EP17}, that the
lattice EH indeed deviates from the CFT prediction, which can only be recovered after taking a proper
continuum limit \cite{ABCH17,ETP19,DGT20,RSC22}. Nevertheless, it has been demonstrated on a number of
examples, that the simple lattice discretization of the CFT ansatz provides an excellent approximation
of the actual EH at low energies and for large subsystems \cite{DVZ18,GMCD18,MGDR19,ZCDR20}.

Here we explore a different scenario, where the model is defined in continuous space, but described
by the nonrelativistic Schr\"odinger equation. We focus on the free-fermion gas, where the entanglement
entropy has been studied before \cite{CMV11,CMV12,MPST22}, and shows a logarithmic area-law violation
in arbitrary dimensions due to the presence of a Fermi surface \cite{GK06,Wolf06,BCS06,LSS14,PS23}. 
Although this result was interpreted via the contributions of independent gapless modes building
up the Fermi surface \cite{Swingle10}, the precise applicability of a CFT description in higher dimensions
remained elusive.

Our main goal here is to directly address the EH of the Fermi gas for a $d$-dimensional spherical domain $A$
with radius $R$, and compare it to the CFT prediction \cite{HL82,CHM11}
\eq{
\ehcft=\frac{\pi R}{v} \int_A \dd^d \mathbf{x} \,
\left(1-\frac{|\mathbf{x}|^2}{R^2}\right) T_{00}(\mathbf{x}),
\label{EHCFT}}
where $T_{00}(\mathbf{x})$ is the energy density and $v$ is the speed of excitations, which makes
$\ehcft$ dimensionless. Its form thus corresponds
to an inverse temperature that varies parabolically in the radius and vanishes at the surface of the sphere.
The numerical check of \eqref{EHCFT} for a free massless scalar field was carried out by first decomposing
the EH into angular momentum sectors, and then discretizing the remaining radial problem \cite{JT22}.
While a good agreement with CFT was found at low angular momenta, for higher ones the results are
inconclusive.

Our main result is that, in any dimension $d>1$, the CFT description of the nonrelativistic Fermi
gas breaks down at large angular momenta. In particular, we show the equivalence of the
entanglement spectra in continuous free space to those of a lattice problem with a \emph{linear}
potential \cite{EIP09}. The mapping identifies the angular momentum with the subsystem
boundary on the chain, whereas the radius $R$ sets the length of the region with nontrivial
fermion density. While the bulk of this region admits an effective CFT description \cite{DSVC17},
characterized by a spatially varying Fermi velocity, the fine structure
close to the dilute edge is not properly captured. The discrepancy is demonstrated
by comparing the actual entanglement spectra and entropies to those that follow from
parabolic deformations \eqref{EHCFT} of the physical Hamiltonian, which
commute exactly with the EH \cite{SP61,Slepian64,EP13}.

The free Fermi gas in $d$ dimensions is described by the single-particle Hamiltonian
\eq{
\hat H = \frac{\mathbf{\hat p}^2}{2m} - \mu ,
\label{H}}
where $\mathbf{\hat p}=-i \nabla$ is the momentum operator and the chemical potential
$\mu=q_F^2/2m$ sets the filling via the Fermi wavenumber $q_F$. The ground state is
given by a Fermi sea $F$, with the plane-wave modes occupied in a spherical domain $|\mathbf{q}|<q_F$.
We are interested in a spherical subsystem $A$ of radius $R$ centered around the origin,
$|\mathbf{x}|<R$. The entanglement Hamiltonian $\mathcal{\hat H}$ is then defined via the
reduced density matrix and can be written as \cite{PE09}
\eq{
\hat \rho_A = \frac{1}{\mathcal{Z}} \ee^{-\mathcal{\hat H}}, \qquad
\mathcal{\hat H}=\ln (\mathcal{\hat K}^{-1}_A-1),
\label{EH}}
in terms of an integral operator
\eq{
(\mathcal{\hat K}_A \, \psi)(\mathbf{x}) = \int_A \dd^d\mathbf{x'} \,
K(\mathbf{x},\mathbf{x'}) \, \psi(\mathbf{x'}) \, ,
\label{KA}}
that acts on wavefunctions in the domain $A$, with the kernel given by the two-point
correlation function
\eq{
K({\bf x},{\bf x'}) = \int_F \frac{\dd^d\mathbf{q}}{(2\pi)^d} \,
\ee^{i {\bf q}({\bf x} -{\bf x'})} \, .
\label{K}}

We first discuss the simplest case of a 1D system, where $A=\left[-R,R\right]$.
After a rescaling $y=x/R$, the integral operator \eqref{KA} is given by the famous sine kernel
\eq{
K(y,y') = \frac{\sin c(y-y')}{\pi (y-y')} ,
\label{sink}}
which depends on the dimensionless parameter $c=q_F R$.
To construct the EH via \eqref{EH}, one needs to solve $\mathcal{\hat K}_A \psi_k = \zeta_k \psi_k$
to find the eigenvalues and eigenfunctions of $\mathcal{\hat K}_A$. This can be done by
considering instead the \emph{differential} operator \cite{Ince56,SP61,Dnote}
\eq{
\hat D = - \frac{\dd}{\dd y} (1-y^2) \frac{\dd}{\dd y} - c^2(1-y^2) \, ,
\label{D}
}
which commutes with the integral operator, $[\mathcal{\hat K}_A,\hat D]=0$.
The bounded solutions of the equation $\hat D \, \psi_k = \chi_k \, \psi_k$ within the domain
$|y|<1$ are known as the angular prolate spheroidal wavefunctions \cite{MS54,Flammer57},
$\psi_k(y)=S_{0k}(c,y)$, and exist for a discrete set of eigenvalues $\chi_k$ with $k=0,1,\dots$ 
The eigenvalues of $\mathcal{\hat K}_A$ then follow from the radial spheroidal wavefunctions as
$\zeta_k= \frac{2c}{\pi}[R_{0k}(c,1)]^2$ \cite{SP61}.

\begin{figure}[t]
\center
\includegraphics[width=\columnwidth]{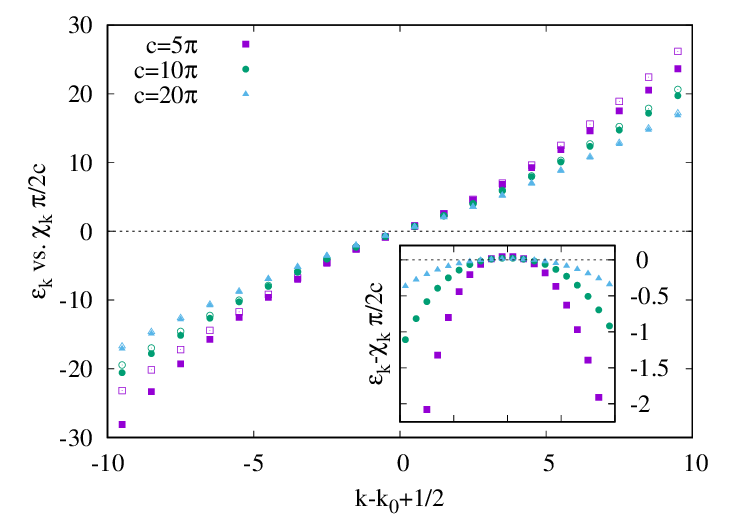}
\caption{Single-particle entanglement spectra $\varepsilon_k$ of $\mathcal{\hat H}$ (full symbols) in one dimension,
compared against the spectra of $\ehcft$ (empty symbols) for various values of $c$. The inset shows the deviations.}
\label{fig:epschi1d}
\end{figure}

It is easy to see, that the operator \eqref{D} is a simple \emph{parabolic deformation} of the
original Hamiltonian \eqref{H}. Comparing with \eqref{EHCFT}, one can identify it with
the CFT expression after proper rescaling
\eq{
\ehcft = \frac{\pi R}{v_F} \frac{\hat D}{2mR^2} = \frac{\pi}{2c} \hat D ,
\label{EHCFT1d}}
where the speed must be identified with the Fermi velocity $v_F=q_F/m$. The spheroidal eigenvalues
$\chi_k$ can be computed using Mathematica, and thus the spectrum of $\ehcft$
can be compared against that $\varepsilon_k=\ln(\zeta_k^{-1}-1)$ of the actual EH in \eqref{EH}.
These are shown in Fig.~\ref{fig:epschi1d}, with the full/empty symbols corresponding to $\varepsilon_k$
and $\frac{\pi}{2c}\chi_k$, respectively, while the inset shows their difference. Note that the
index $k$ was shifted by $k_0-1/2$, with $k_0=2c/\pi$, to align the low-energy part of the spectra.
One clearly observes that the deviation diminishes for increasing $c$, suggesting the asymptotic equivalence
$\mathcal{\hat H} \to \ehcft$ of the operators. This is supported by analytical results \cite{Slepian65,SS65,dCM72},
as well as further numerical evidence \cite{sm}. In particular, for finite $c$ one has a series expansion
\eq{
\mathcal{\hat H} =
\ehcft + \sum_{n=1}^{\infty} \frac{1}{c^n} P_{n+1}(\ehcft),
\label{CFTcorr}}
where $P_n$ is an $n$-th order polynomial. The $n$-th correction term is thus an increasingly non-local
differential operator of order $2(n+1)$, which is, however, suppressed by $c^n$. Furthermore, using the 
lowest order terms in \eqref{CFTcorr}, we find that the entanglement entropy $S=-\Tr[\rho_A \ln \rho_A]$
is reproduced by $\ehcft$ up to a correction scaling as $\delta S \propto \ln (c)/c^2$, which agrees
well with our numerics \cite{sm}.
%

%
%

We now proceed to the case $d \ge 2$, which considerably simplifies using the rotational symmetry of both $A$ and $F$.
Indeed, setting $\mathbf{x}=r\,\mathbf{n}$, the Hamiltonian can be decomposed by considering the ansatz for the wavefunction
\eq{
\psi(r \, \mathbf{n})=\frac{\Phi(r)}{r^{(d-1)/2}}Y_{\ell,i}(\mathbf{n}),
\label{sph}}
where $Y_{\ell,i}(\mathbf{n})$ are $d$-dimensional spherical harmonics \cite{FE12}, with $\mathbf{n}$ being a vector 
on the surface of the unit sphere, parametrized by $d-1$ angular coordinates. The quantum number $\ell=0,1,\dots$ corresponds to
the angular momentum, and $i=1,\dots,M_\ell$ indexes the linearly independent spherical harmonics with fixed $\ell$.
In this basis, the Hamiltonian $\hat H = \bigoplus_{\ell,i} \hat H_{\ell,i}$
becomes block-diagonal and in the respective sector reads
\eq{
\hat H_{\ell,i} =
\frac{1}{2m}\Big(-\frac{\dd^2}{\dd r^2} - q_F^2 + \frac{(\ell+\frac{d-2}{2})^2-1/4}{r^2}  \Big).
\label{Hl}}
Note that $\hat H_{\ell,i}$ does not depend on the quantum number $i$,
such that one simply has a degeneracy in each sector $\ell\ge1$ with corresponding multiplicity
\eq{
M_\ell = \frac{2\ell+d-2}{\ell} \binom{\ell+d-3}{\ell-1},
\label{Ml}}
while $M_0 = 1$. Thus the problem boils down to treating the one-dimensional
Hamiltonian \eqref{Hl}, where one has an extra contribution from the centrifugal potential.
The dimensionality enters via the multiplicities \eqref{Ml} and a shift of the angular momentum
index $\ell$. For simplicity, we shall discuss the 2D case below, as the generalization to $d>2$ is trivial.

The eigenvalue problem of the kernel \eqref{K} was considered in \cite{Slepian64}, see \cite{sm} for details.
One first rewrites it as the absolute square of an exponential kernel $K'({\bf y} ,{\bf z}) = \ee^{i c{\bf y}{\bf z}}$
in the scaled coordinates $\mathbf{y}=\mathbf{x}/R$ and $\mathbf{z}=\mathbf{q}/q_F$.
Separating variables using the ansatz \eqref{sph}, one is led to consider the radial eigenvalue problem
$\mathcal{\hat K'}_{\ell} \Phi_{\ell,k} = \gamma_{\ell,k}\Phi_{\ell,k}$,
with the kernel given by $K'_{\ell}(y,z) =  J_{\ell}(cyz) \sqrt{c^2yz}$. Note that $y=|\mathbf{y}|\le1$, $z=|\mathbf{z}|\le1$,
and the eigenvalues of the original operator $\mathcal{\hat K}_\ell$ follow as
$\zeta_{\ell,k}=|\gamma_{\ell,k}|^2$. The squared kernel can then be written as
\eq{K_\ell(y,y') = 
2c^2 \sqrt{yy'} K_{\mathrm{Be},\ell}(c^2y^2,c^2y'^2)
\label{Kl}}
via the Bessel kernel defined as \cite{TW94b}
\eq{
K_{\mathrm{Be},\ell}(u,v) =
\frac{\sqrt{v} J_\ell(\sqrt{u}) J'_\ell(\sqrt{v})-\sqrt{u} J_\ell(\sqrt{v}) J'_\ell(\sqrt{u})}{2(u-v)}.
\label{KBe}}
Note that the factor in \eqref{Kl} in front of the Bessel kernel can be absorbed by a change
of variables $u=c^2y^2$ and $v=c^2y'^2$, such that the spectrum of $\mathcal{\hat K}_\ell$
on the domain $[0,1]$ is identical to that of $\mathcal{\hat K}_{\mathrm{Be},\ell}$
on $[0,c^2]$.

Analogously to the 1D case, one can find again a commuting differential operator
in each angular momentum sector, $[\mathcal{\hat K}_\ell,\hat D_\ell]=0$, which reads
\cite{Slepian64,Dlnote}
\eq{
\hat D_\ell = -\frac{\dd}{\dd y} \beta(y) \frac{\dd}{\dd y} - \Big(c^2 - \frac{\ell^2-1/4}{y^2}\Big)\beta(y)\, ,
\label{Dl}
}
with $\beta(y)=1-y^2$. Clearly, \eqref{Dl} can be interpreted as the parabolic deformation of
the radial Hamiltonian \eqref{Hl}. Its eigenvalue equation reads
$\hat D_\ell \Phi_{\ell,k} = \chi_{\ell,k} \Phi_{\ell,k}$, and the
eigenfunctions were dubbed generalized prolate spheroidal wavefunctions. Their asymptotic
expressions for $c,k \gg1 $ were studied in \cite{Slepian64}. Moreover, high precision numerical computation
of the eigenvalues $\zeta_{\ell,k}$ and $\chi_{\ell,k}$ is available via an open-source MATLAB code
\cite{Lederman17}.

Before turning to the numerics, however, one needs an argument to fix the velocity in the CFT expression
\eqref{EHCFT}. Indeed, the inhomogeneous part of the radial Hamiltonian \eqref{Hl} can be interpreted
as a spatially varying chemical potential $\mu_\ell(r)$. In other words, the effective Fermi energy of the
radial motion is reduced by the centrifugal energy of the orbital one. Furthermore, we argue that the only
relevant radius in our problem is that of our subsystem, and thus the effective chemical potential should be
evaluated at $r=R$. Assuming $R \gg 1$, one obtains for the radial Fermi velocity
\eq{
v_{F,\ell} = \sqrt{\frac{2\mu_\ell(R)}{m}}=
v_F\sqrt{1-\frac{\ell^2}{c^2}} .
\label{vFl}}
In particular, $v_{F,\ell}$ vanishes at $\ell = c$, which corresponds to the angular momentum
where the classical turning point is given by $R$. For all $\ell>c$, the eigenfunctions of \eqref{Hl}
have exponentially small amplitudes within $A$, and thus their contribution to the EH should be
negligible.

Alternatively, the emergence of the Fermi velocity \eqref{vFl} can be understood by mapping the
problem to that of an inhomogeneous quantum chain. This can be achieved using a remarkable
identity found in Ref.~\cite{MNS19}, which establishes a connection between the
Bessel kernel \eqref{KBe} and the analogous discrete Bessel kernel
\eq{
K_{\mathrm{dBe},c}(i,j)=
\frac{c \, J_{i-1}(c) J_{j}(c) - c\, J_{i}(c) J_{j-1}(c)}{2(i-j)},
\label{KdBe}}
where $i,j \in \mathbb{Z}$. The identity relates the trace of an integer power
of the corresponding operators \cite{MNS19}
\eq{
\Tr_{[0,c^2]}(\mathcal{\hat K}^n_{\mathrm{Be,\ell}})=
\Tr_{[\ell+1,\infty)}(\mathcal{\hat K}^n_{\mathrm{dBe},c}) \, ,
\label{TrB}}
where the subscripts denote the domains of the respective kernels, over which
the trace is carried out, with the r.h.s. being the trace of an ordinary matrix.
Since the relation holds for arbitrary $n$, this implies
that the spectra of the two operators are identical.

The matrix defined in \eqref{KdBe} is precisely the correlation matrix
of a hopping chain with a \emph{linear} potential \cite{EIP09},
and unitary equivalent to the one describing domain-wall melting \cite{SK23}.
The parameter $c$ now plays the role of the half-width of the front region, where
the fermion density differs from one and zero. Moreover, the angular momentum
$\ell$ is identified with the position of the entanglement cut. In turn, the expression
\eqref{vFl} simply corresponds to the spatial dependence of the Fermi velocity due to
the variation of the filling within the front region \cite{ADSV16,EB17}.
The CFT prediction for the respective EH thus reads
\eq{
\ehlcft = \frac{\pi}{2\sqrt{c^2-\ell^2}} \hat D_\ell \, .
\label{EHlCFT}}

\begin{figure}[t]
\center
\includegraphics[width=0.49\textwidth]{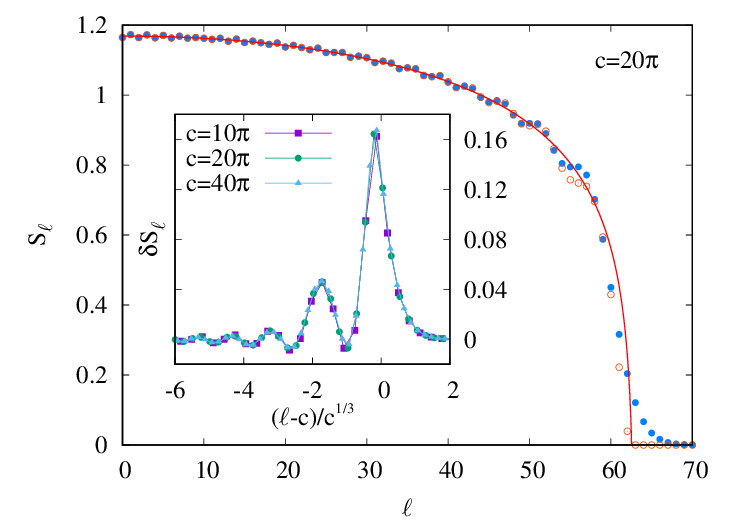}
\caption{Entanglement entropies $S_\ell$ (full symbols) and $S_{\ell,\mathrm{\tiny{CFT}}}$ (empty symbols),
calculated using the ansatz \eqref{EHlCFT} for $c=20\pi$. The red solid line shows the result \eqref{Sl}.
Inset: difference of the entropies in the appropriately rescaled edge regime, for various values of $c$.}
\label{fig:sl2d}
\end{figure}

To test the validity of the ansatz \eqref{EHlCFT}, we evaluate and compare the entropies
obtained from $\mathcal{\hat H}_\ell$ and $\ehlcft$,
as shown in Fig.~\ref{fig:sl2d}. The agreement is excellent in the bulk of the profile, where
the asymptotics of the spectra $\elk$ with $\ell/c$ fixed were studied numerically for the
gradient chain \cite{EP14}. The resulting entropy profile
\eq{
S_\ell = \frac{1}{6}\ln(c) + \frac{1}{4}\ln[1-(\ell/c)^2]+ \mathcal{C} ,
\label{Sl}}
where $\mathcal{C}\approx0.4785$ is a nonuniversal constant \cite{JK04,SI12}, is shown by the red line.
In fact, \eqref{Sl} can also be derived using a curved-space CFT approach \cite{DSVC17},
where the inhomogeneous metric is chosen to absorb the spatial variation of the Fermi velocity.
While \eqref{Sl} gives an accurate description of the bulk entropy profile, it does not capture
the fine structure around the edge $\ell \approx c$, where also the ansatz \eqref{EHlCFT} seems to 
break down. Indeed, using the scaling variable $(\ell-c)/c^{1/3}$, the correlation matrix \eqref{KdBe}
can be approximated by the Airy kernel \cite{TW94a}, and $S_\ell$ displays a corresponding edge
scaling \cite{ER13}. As shown by the inset of Fig.~\ref{fig:sl2d}, the same holds true for the difference
$\delta S_{\ell}=S_{\ell}-S_{\ell,\mathrm{\tiny{CFT}}}$, which shows a data collapse for various values of $c$. 

%
%

\begin{figure}[t]
\center
\includegraphics[width=0.49\textwidth]{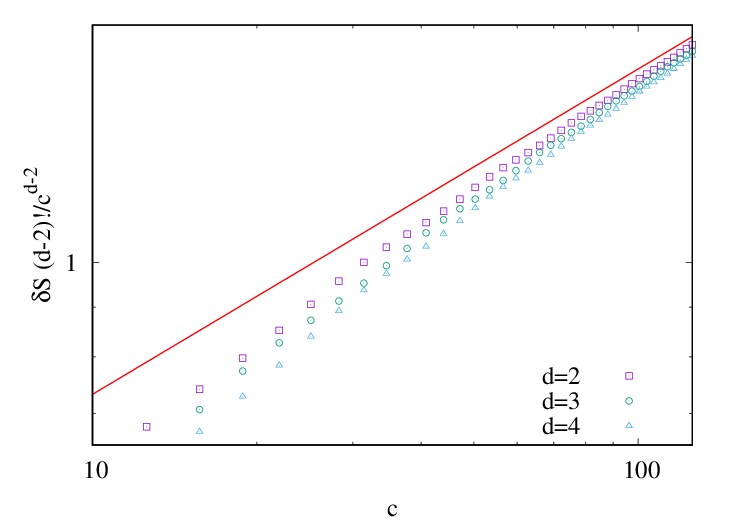}
\caption{Deviation of the total entropy from the one calculated via the CFT EH for various dimensions,
scaled according to \eqref{dSdim}. The red solid line with slope $1/3$ is a guide to the eye.}
\label{fig:dSdim}
\end{figure}

The situation is very similar in $d>2$ dimensions, where the index of the Bessel kernel is
$\ell + (d-2)/2$. This is a half-integer in odd dimensions, such that the one-to-one correspondence
with the gradient chain is lost. Nevertheless, when plotted against the shifted index $\ell + (d-2)/2$,
the entropy profile $S_\ell$ smoothly interpolates between the data points of the $d=2$ case.
Applying the same shift in the scaling factor in \eqref{EHlCFT}, the plot of the 3D case is almost
identical to Fig.~\ref{fig:sl2d}. One thus concludes that the CFT ansatz breaks down for high angular
momenta $\ell \approx c-(d-2)/2$. Due to the increasing multiplicities $M_\ell$ with the
dimensionality, however, the leading order mismatch of the total entropy in $d\ge2$ scales as
\eq{
\delta S =\sum_{\ell} M_\ell \, \delta S_\ell
\propto \frac{c^{d-2}}{(d-2)!} c^{1/3} \, .
\label{dSdim}}
Thus, in sharp contrast to the 1D case, the entropy deviation becomes divergent in
the $c\to\infty$ limit. This is a consequence of the edge-scaling regime in angular-momentum space,
which is not properly described by CFT. The scaling \eqref{dSdim} is consistent with our numerics
in Fig.~\ref{fig:dSdim}, albeit with strong subleading corrections.

%
%

The mapping to the gradient chain, with resulting entropy profile \eqref{Sl}, also allows
us to obtain the analytical result for the total entropy
\eq{
S =\sum_{\ell} M_\ell \, S_\ell
\simeq \sigma_d \, c^{d-1} \ln c + A_d \, c^{d-1},
\label{Stot}}
where the prefactors can be calculated as \cite{sm}
\eq{
\sigma_d =
\frac{1}{3(d-1)!}, \qquad
A_d = \frac{4 \, \mathcal{C} - \psi(\frac{d+1}{2}) - \gamma}{2(d-1)!},
\label{sigdAd}}
with $\psi(x)$ being the digamma function and $\gamma$ the Euler-Mascheroni constant.
It is easy to check that the prefactor of the area-law violating term agrees with the general
expression found in \cite{GK06,LSS14}. The area-law contribution is nonuniversal, and
follows from the summation of the second and third terms in \eqref{Sl} \cite{sm}.
We tested the prediction \eqref{Stot} by adding a subleading term $B_d \, c^{d-2}$ and
fitting to our numerical data. The results $\sigma_2=0.3332$, $A_2=0.651$ and
$\sigma_3=0.1667$, $A_3=0.2285$ for the 2D and 3D cases, respectively, are in excellent
agreement with \eqref{sigdAd}. Note that our result on $A_d$ also agrees with the conjecture
formulated in \cite{SDMS21}. One should also remark that, in free massless relativistic theories,
no violation of the area law occurs \cite{Srednicki93,HV23a,HV23b}.

In conclusion, we have found that the EH of a nonrelativistic Fermi gas is well reproduced
by the appropriately rescaled parabolic deformation of the physical Hamiltonian. While in 1D
the relation becomes asymptotically exact in the limit of large subsystems, the situation in
higher dimensions is much more subtle. Firstly, the CFT prediction can only be applied
in the angular momentum sectors, after rescaling with the Fermi velocity of the radial motion.
Since this velocity carries a nontrivial dependence on $\ell$, the relation cannot be lifted back
to the total EH and rewritten as a deformation of the total energy density as in \eqref{EHCFT}.
In sharp contrast, for relativistic Dirac fermions the form of the total EH is identical to those
in the sectors \cite{HV23b}. Secondly, for the nonrelativistic case some deviations persist 
even in the sectors for large angular momenta. Using the mapping to the gradient chain,
these discrepancies can be traced back to the dilute edge regime of the fermionic density,
where the fine-structure of the correlations does not admit a CFT description. Hence,
translating \eqref{EHCFT} to nonrelativistic systems requires proper insight and care.

Our work opens up various directions for future research. One could address how the shape
of the Fermi surface, which is known to be crucial for the entropy scaling \cite{GK06,LSS14},
affects the results for the EH. A further natural extension would be the study of a trapped
Fermi gas \cite{DDMS19}, where the CFT predictions for the EH are also available \cite{TRS18}.
Finally, one should investigate how the results generalize to particles with bosonic statistics.

We thank I. Peschel and E. Tonni for fruitful discussions and correspondence. In our numerical calculations
we used the open-source code available under \url{http://github.com/lederman/prol}.
The author acknowledges funding from the Austrian Science Fund (FWF) through project No. P35434-N.

\vspace{-5mm}

\bibliography{ehfermigas_refs}

\begin{thebibliography}{68}
\bibitem [{Dno()}]{Dnote}%
  \BibitemOpen
  \href@noop {} {}\bibinfo {note} {Note that the differential operator used in
  Ref. [36] differs by an additive constant $c^2$.}\BibitemShut {Stop}%
\bibitem [{Dln()}]{Dlnote}%
  \BibitemOpen
  \href@noop {} {}\bibinfo {note} {Note that the differential operator used in
  Ref. [37] differs by an additive constant $c^2+\ell^2-1/4$.}\BibitemShut
  {Stop}%
\end{thebibliography}

\onecolumngrid
\clearpage

\begin{center}
\textbf{\large Supplemental Material: Entanglement Hamiltonian of a nonrelativistic Fermi gas}
\end{center}
\author{Viktor Eisler}
\affiliation{
Institute of Theoretical and Computational Physics,
Graz University of Technology, Petersgasse 16, 8010 Graz, Austria}

\setcounter{equation}{0}
\setcounter{figure}{0}
\setcounter{table}{0}
\setcounter{page}{1}
\makeatletter
\renewcommand{\theequation}{S\arabic{equation}}
\renewcommand{\thefigure}{S\arabic{figure}}


\section{Corrections to CFT result in 1D}

In the following we present some results on the first corrections to the CFT form of the
spectrum in \eqref{CFTcorr}. The problem amounts to finding an asymptotic relation between
the eigenvalues $\varepsilon_k$ and $\chi_k$, as a power series expansion in terms of $1/c$.
Introducing $b=\varepsilon_k/\pi$ and dropping the eigenvalue index $k$, the first few terms
of such an expansion were provided in \cite{Slepian65,SS65} as
\eq{
\chi = 2bc + \frac{b^2-1}{2} - \frac{b^3-b}{8c}+\mathcal{O}(c^{-2}) \, .
\label{chieps}}
Note that our convention differs from the one in \cite{Slepian65,SS65} in the definition of the differential
operator by a constant shift $c^2$, which is subtracted in our case. Then the leading order
is linear in $c$ and gives just the CFT result \eqref{EHCFT1d}. Inverting the relation \eqref{chieps}
to lowest order in $1/c$ one finds
\eq{
b=\frac{\varepsilon}{\pi} =
\frac{\chi}{2c} + \frac{1-(\frac{\chi}{2c})^2}{4c}
+\mathcal{O}(c^{-2}) \, .
\label{epschi_series}}
One thus obtains a perturbation series in terms of $1/c$, where each term is a polynomial of the
variable $\frac{\chi}{2c}$.

\begin{figure}[htb]
\center
\includegraphics[width=0.49\textwidth]{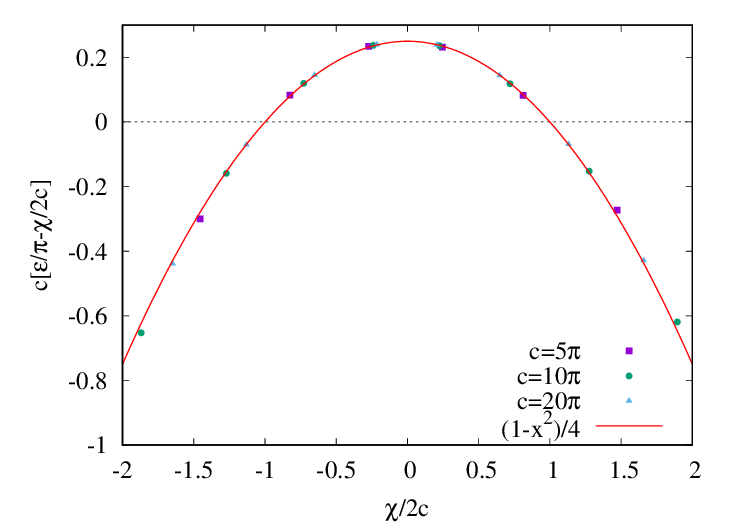}
\includegraphics[width=0.49\textwidth]{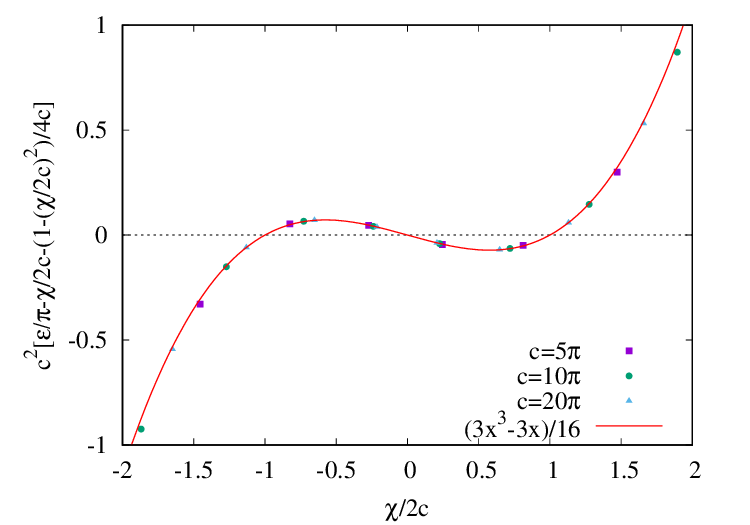}
\caption{Corrections of order $c^{-1}$ (left) and $c^{-2}$ (right) to the CFT ansatz of the spectrum.}
\label{fig:epschi_series}
\end{figure}

The result can be checked numerically by subtracting the leading (CFT) term and plotting 
the corrections rescaled by $c$ as a function of $x=\frac{\chi}{2c}$. This is shown on the left of
Fig.~\ref{fig:epschi_series} for various values of $c$, and compared against the analytic result
depicted by the red line. The agreement is very good for small values of $\frac{\chi}{2c}$,
but deviations are already visible for larger values. Moreover, one can also probe the
$1/c^2$ correction by subtracting the $1/c$ term and rescaling the difference by $c^2$,
as shown on the right of Fig.~\ref{fig:epschi_series}. One has again a good scaling collapse,
and our fit for small parameter values $x$ suggests the third-order polynomial $P_{3}(x)=(3x^3-3x)/16$.
We note that higher order corrections could, in principle, be studied systematically by
using the results of Ref. \cite{dCM72}, where uniform asymptotic expressions for both $\chi$ and
$\varepsilon$ eigenvalues were obtained via the WKB method.

%
%

We now move to study the corrections in the entanglement entropy $S$, which has the thermal form
\eq{
S = \sum_{k=0}^{\infty} s(\varepsilon_k) \, , \qquad
s(\varepsilon) = \frac{\varepsilon}{\ee^{\varepsilon}+1} +
\ln (1+\ee^{-\varepsilon}) \, ,
\label{S}}
in terms of the eigenvalues $\varepsilon_k$ of the EH. Although one has an infinite sum, the dominant
contribution comes from the low-lying entanglement spectrum. The asymptotic expansion, valid for $c \gg 1$
and large indices $k$ around the center of the spectrum $k_0=2c/\pi$, is given by \cite{Slepian65}
\eq{
\frac{\varepsilon_k}{2\pi} \ln(4c)-\varphi \Big(\frac{\varepsilon_k}{2\pi}\Big) =
\frac{\pi}{2} \Big(k-k_0+\frac{1}{2}\Big) \, ,
\label{epsslep}
}
where $\varphi(z)=\mathrm{arg}\, \Gamma(1/2+iz)$ and $\Gamma(z)$ is the gamma function.
The result for the entropy can be found by introducing the density of states
$\rho(\varepsilon) = \frac{\dd k}{\dd \varepsilon}$ and turning the sum in \eqref{S} into an integral
\eq{
S = \int \dd \varepsilon \, \rho(\varepsilon) \, s(\varepsilon) =
\frac{1}{3} \ln (4c) + S_0 \, .
\label{Sint}}
The leading logarithmic scaling with argument $4c=4q_FR$ comes from the low-lying linear
regime of the spectrum, whereas the nonuniversal constant
\eq{
S_0 = -\frac{1}{\pi^2}\int_{-\infty}^{\infty} \dd \varepsilon \, s(\varepsilon)
\varphi'\left(\frac{\varepsilon}{2\pi}\right),
\label{S0}}
is due to the curvature and has a numerical value $S_0 \approx 0.495$ \cite{JK04,SI12}.

We now calculate the entropy from the CFT ansatz \eqref{EHCFT1d}, which gives
\eq{
S_{\mathrm{\tiny{CFT}}} = \sum_{k=0}^{\infty} s \Big(\frac{\pi\chi_k}{2c}\Big) \approx
\int_{-\infty}^{\infty} \dd \varepsilon \, \rho(\varepsilon)
\Big[ s(\varepsilon) + s'(\varepsilon) \, \delta(\varepsilon) +  \frac{1}{2} s''(\varepsilon) \, \delta^2(\varepsilon) \Big] ,
\label{SCFT}}
where $\delta(\varepsilon)$ denotes the correction term in \eqref{chieps}. Note that
$\rho(\varepsilon)$ and $s''(\varepsilon)$ are even functions, whereas $s'(\varepsilon)$ is odd.
Hence, the lowest nonvanishing contribution comes from the odd part of $\delta(\varepsilon)$
and the even part of $\delta^2(\varepsilon)$, respectively. Furthermore, the dominant contribution
comes from the constant part $\rho(\varepsilon)\approx\ln(4c)/\pi^2$ of the spectral density.
It turns out that the corresponding integrals in \eqref{SCFT} can be evaluated explicitly, and one arrives at
\eq{
\delta S = S-S_{\mathrm{\tiny{CFT}}}=-\frac{\ln(c)}{40 \, c^2} + \mathcal{O}(c^{-2}) \, .
\label{dS1d}}
%

%
\begin{figure}[htb]
\center
\includegraphics[width=0.49\textwidth]{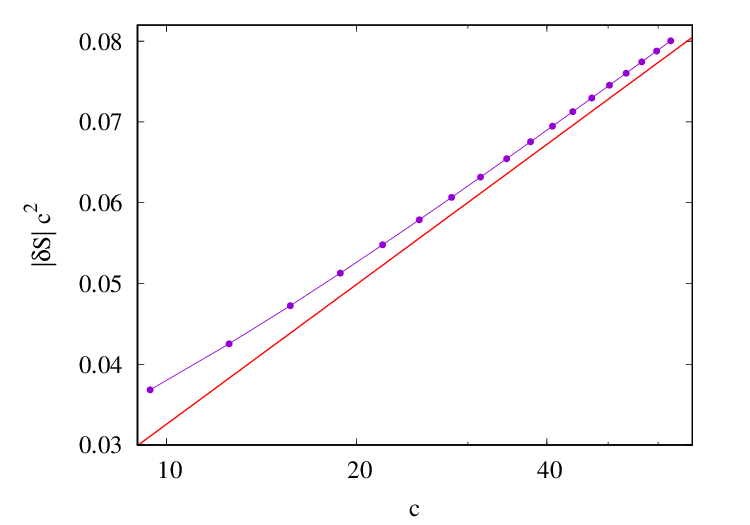}
\caption{Deviation of the entropy from the CFT result, scaled by $c^2$ and plotted on a horizontal
logarithmic scale. The red solid line has slope $1/40$ according to \eqref{dS1d}.}
\label{fig:dS1d}
\end{figure}
%

We conclude this section by some comments on the higher dimensional case.
In the limit $k, c \gg 1$ and $\ell$ fixed, the asymptotics of the eigenvalues $\elk$ and $\clk$
were considered in \cite{Slepian64}. However, we need instead the regime $\ell \sim c$,
where the spectrum was studied in the equivalent lattice problem of the gradient chain \cite{EP14}.
In particular, one finds
\eq{
\frac{\elk}{2\pi} \ln \Big[4c\Big(1-\frac{\ell^2}{c^2}\Big)^{3/2}\Big]
-\varphi \Big(\frac{\elk}{2\pi}\Big) =
\pi \Big(k -k_0 +\frac{1}{2}\Big),
\label{eps2d}
}
where $k_0$ is given by the expectation value of the particle number within the subsystem. In fact,
this result was found by fitting the spectral function $k(\varepsilon)$ to a modified 1D ansatz \eqref{epsslep},
with the argument of the logarithm and the offset $k_0$ being fit parameters. Note also the change
of the factor $\pi/2 \to \pi$ on the r.h.s. of \eqref{eps2d}, which leads to a halved prefactor of the logarithmic
term in the entropy \eqref{Sl}, corresponding to only one boundary point of the subsystem. The constant 
term in \eqref{Sl} is related to \eqref{S0} as $\mathcal{C}=S_0/2+\ln(2)/3$. It should be stressed, however,
that \eqref{eps2d} is not expected to work properly in the edge regime $|\ell - c| \propto c^{1/3}$, where
also the CFT ansatz \eqref{EHlCFT} breaks down.

\section{Kernel for $d \ge 2$}

Let us rewrite the correlation kernel \eqref{K}, with spatial $\mathbf{y}=\mathbf{x}/R$ and momentum
$\mathbf{z}=\mathbf{q}/q_F$ variables rescaled to the unit sphere, as the square of another kernel
\eq{
K({\bf y},{\bf y'}) = \Big(\frac{c}{2\pi}\Big)^d
\int_{|{\bf z}|<1} \dd^d z \, K'({\bf y} ,{\bf z})\bar K'({\bf z} ,{\bf y'}),
\label{K2}}
where $K'({\bf y} ,{\bf z}) = \ee^{i c{\bf y}{\bf z}}$ and the bar denotes complex conjugation.
Due to rotational symmetry, the eigenvalue problem of the corresponding integral operator
can be decoupled in angular momentum sectors by introducing the ansatz \eqref{sph}.
Setting $\mathbf{y} = r \, \mathbf{n}$ and $\mathbf{z} = r' \, \mathbf{n'}$, with $\mathbf{n}$ and
$\mathbf{n'}$ being vectors on the surface of the $d$-dimensional unit sphere, the integral operator acts as
\eq{
(\mathcal{\hat K'} \, \psi)(r \, \mathbf{n}) = 
\int_{0}^{1} \dd r' \, r'^{d-1} \frac{\Phi(r')}{r'^{(d-1)/2}}
\int_{\Omega'} \dd \Omega' \, \ee^{i c r r' {\bf n}{\bf n'}} \, Y_{\ell,i}(\mathbf{n'}) \, .
\label{Kp}
}
It was shown in Ref. \cite{Slepian64} that the second integral on the surface $\Omega'$ of the unit sphere
can be carried out as
\eq{
\int_{\Omega'} \dd \Omega' \, \ee^{i c r t {\bf n}{\bf n'}} \, Y_{\ell,i}(\mathbf{n'})=
i^{\ell} (2\pi)^{d/2}\frac{J_{\ell+\frac{d-2}{2}}(crr')}{(crr')^{\frac{d-2}{2}}} Y_{\ell,i}(\mathbf{n}) \, .
}
Hence the eigenvalue problem in the sector $(\ell,i)$ reads
\eq{
(\mathcal{\hat K'}_\ell \, \Phi_k)(r) =
\int_{0}^{1} \dd r' K'_{\ell}(r,r') \, \Phi_k(r') =
\gamma_{\ell,k} \, \Phi_k(r) \, ,
}
with the kernel given by
\eq{
K'_{\ell}(r,r') =  J_{\ell+\frac{d-2}{2}}(crr') \sqrt{c^2rr'} .
\label{Kpl}}
The kernel does not depend explicitly on the index $i$ of the spherical harmonics, and thus
just gives a multiplicity $M_\ell$ in the corresponding sector, similarly to the case of the physical Hamiltonian.
Note that, for simplicity, we omitted a constant factor from the definition of the kernel \eqref{Kpl}, which should
be kept track of in the integral operator
\eq{
\mathcal{\hat K'} = \bigoplus_{\ell,i} i^{\ell} \Big(\frac{2\pi}{c}\Big)^{d/2} \mathcal{\hat K'}_\ell \, .
\label{Kpsum}}
However, when taking the square of the kernel in \eqref{K2}, the prefactor in front of the integral
exactly cancels the one included in \eqref{Kpsum}, and one has
$\mathcal{\hat K} = \bigoplus_{\ell,i} \mathcal{\hat K}_\ell$ with underlying kernel
\eq{
\qquad
K_\ell(r,r') = \int_{0}^{1} \dd t \, K_\ell(r,t) \, K_\ell(t,r') \, .
\label{Kl2}}

Finally, the kernel $K_\ell(r,r')$ can be related to the Bessel kernel, which has the integral representation \cite{TW94b}
\eq{
K_{\mathrm{Be},\alpha}(x,y) =
\frac{1}{4}\int_{0}^{1} \dd z \, J_\alpha(\sqrt{xz}) \, J_\alpha(\sqrt{yz}) \, .
\label{Kr}}
Indeed, inserting \eqref{Kpl} into \eqref{Kl2} and changing variables as $z=t^2$ with $\dd z  = 2 t \, \dd t$,
one arrives at the expression
\eq{K_\ell(r,r') = 
2c^2 \sqrt{rr'} K_{\mathrm{Be},\ell+\frac{d-2}{2}}(c^2r^2,c^2r'^2) \, .
}
Furthermore, the trace of an integer power  can be written as
\eq{
\Tr_{[0,1]} (\mathcal{\hat K}^n_\ell)=
\int_{0}^{1} \dd r_1 \dots \dd r_n \,
\prod_{j=1}^{n} 2c^2 r_j K_{\mathrm{Be},\ell}(c^2r^2_j,c^2r^2_{j+1}) =
\int_{0}^{c^2} \dd x_1 \dots \dd x_n \,
\prod_{j=1}^{n} K_{\mathrm{Be},\ell}(x_j,x_{j+1}) \, ,
}
where $r_{n+1}=r_1$, and we have changed variables $x_j = c^2r_j^2$ with $\dd x_j = 2c^2 r_j \, \dd r_j$.
One thus finds
\eq{
\Tr_{[0,1]} (\mathcal{\hat K}^n_\ell)= \Tr_{[0,c^2]} (\mathcal{\hat K}^n_{\mathrm{Be},\ell}) \, .
}

\section{Commuting differential operators}

We shall prove here that the differential operator $\hat D$ defined in \eqref{D} indeed commutes
with the integral operator $\mathcal{\hat K}$ with the sine kernel \eqref{sink} in 1D.
Setting $\beta(x)=1-x^2$, one has
\begin{align}
(\hat D \mathcal{\hat K} f)(x) =
-\Big[\beta(x)\frac{\dd^2}{\dd x^2}+\beta'(x)\frac{\dd}{\dd x}+c^2 \beta(x)\Big]
\int_{-1}^{1} dy \, K(x-y) f(y) \nonumber \\
= -\int_{-1}^{1} dy \, f(y) \Big[\beta(x)\frac{\dd^2}{\dd y^2}-\beta'(x)\frac{\dd}{\dd y}+c^2\beta(x) \Big] K(x-y).
\label{DK}
\end{align}
On the other hand, exchanging the order of the operators one obtains
\begin{align}
(\mathcal{\hat K} \hat D f)(x) 
&= -\int_{-1}^{1} dy \, K(x-y)
\Big[\beta(y)\frac{\dd^2}{\dd y^2}+\beta'(y)\frac{\dd}{\dd y}+c^2 \beta(y)\Big] f(y) \nonumber \\
&=-\int_{-1}^{1} dy \, f(y)
\Big[\frac{\dd^2}{\dd y^2}\beta(y)-\frac{\dd}{\dd y}\beta'(y)+c^2 \beta(y)\Big] K(x-y) \nonumber \\
&=-\int_{-1}^{1} dy \, f(y)
\Big[\beta(y)\frac{\dd^2}{\dd y^2}+\beta'(y)\frac{\dd}{\dd y}+c^2 \beta(y)\Big] K(x-y) .
\label{KD}
\end{align}
In the second line we have integrated by parts, with the derivatives acting on all the functions to their right.
Note that the boundary contributions to the partial integral vanish due to $\beta(\pm1)=0$.
In the third line we have expanded the derivatives using the chain rule. Collecting the terms in
\eqref{DK} and \eqref{KD}, the commutator reads
\eq{
([\hat D , \mathcal{\hat K}] f)(x) 
=\int_{-1}^{1} dy \, f(y) \left\{
(\beta(y)-\beta(x)) \left[\frac{\dd^2}{\dd y^2}K(x-y)+c^2 K(x-y) \right] +
(\beta'(x)+\beta'(y))\frac{\dd}{\dd y}K(x-y)  \right\}.
}
The derivatives of the sine-kernel can be evaluated as
\begin{align}
&\frac{\dd}{\dd y}K(x-y) = \frac{\sin c(x-y)}{\pi (x-y)^2} - c\frac{\cos c(x-y)}{\pi (x-y)}, \\
&\frac{\dd^2}{\dd y^2}K(x-y) = -c^2\frac{\sin c(x-y)}{\pi (x-y)} - 2c\frac{\cos c(x-y)}{\pi (x-y)^2}+2\frac{\sin c(x-y)}{\pi (x-y)^3}.
\end{align}
Using $\beta(y)-\beta(x)=x^2-y^2$ and $\beta'(x)+\beta'(y)=-2(x+y)$, it is easy to check that the commutator indeed vanishes.

The calculation of the commutator for the radial kernel follows a similar route, including the extra centrifugal potential
in the differential operator \eqref{Dl}. The weight function $\beta(x)$ is unchanged, however, one has now a kernel
which depends on the product (instead of the difference) of its variables, $K_\alpha(xy)=J_{\alpha}(cxy)\sqrt{c^2xy}$ with
$\alpha=\ell+(d-2)/2$. This changes the result on the second line of \eqref{DK}, since one has the relation
\eq{
x^n\frac{\dd^n}{\dd x^n} K(xy) = y^n\frac{\dd^n}{\dd y^n} K(xy) \, .
}
Using the parabolic form $\beta(x)=1-x^2$, one can thus write
\begin{align}
(\hat D_\alpha \mathcal{\hat K}_\alpha f)(x)
&=-\int_{0}^{1} dy \, f(y) \Big[\Big(\frac{y^2}{x^2}-y^2\Big)\frac{\dd^2}{\dd y^2}-2y\frac{\dd}{\dd y}+
\Big(c^2-\frac{\alpha^2-1/4}{x^2}\Big)(1-x^2) \Big] K_\alpha(xy) \, , \label{DK2} \\
(\mathcal{\hat K}_\alpha \hat D_\alpha f)(x)
&=-\int_{0}^{1} dy \, f(y)
\Big[(1-y^2)\frac{\dd^2}{\dd y^2}-2y\frac{\dd}{\dd y}+
\Big(c^2-\frac{\alpha^2-1/4}{y^2}\Big)(1-y^2) \Big] K_\alpha(xy) \, .
\label{KD2}
\end{align}
Note that the limits of integration have changed, as one is dealing with a radial problem now.
Since $\beta(0)\ne0$, one has to require that the function vanishes at the origin, $f(0)=0$, to cancel
the boundary contributions in the partial integration. The first derivatives then cancel in the commutator,
and the second derivative can be rewritten as $\frac{\dd^2}{\dd y^2} K_\alpha(xy)=x^2K''_\alpha(xy)$,
where the prime denotes the derivative w.r.t. the argument $xy$. In turn one has
\eq{
([\hat D_\alpha, \mathcal{\hat K}_\alpha] f)(x)=
\int_{0}^{1} dy \, f(y) \, (x^2-y^2)
\Big[ K''_\alpha(xy) + \Big(c^2-\frac{\alpha^2-1/4}{x^2y^2}\Big)K_\alpha(xy) \Big] \, .
\label{comm2}}

Finally, it is easy to show that the expression in the square brackets in \eqref{comm2} vanishes.
Indeed, introducing the variable $z=cxy$ and writing $K_\alpha(xy)=J_\alpha(z)\sqrt{cz}$ one has
\eq{
K''_\alpha(xy) = c^{5/2}\frac{\dd^2}{\dd z^2} \big[ J_\alpha(z) \sqrt{z} \big]=
c^{5/2}z^{-3/2}\Big[z^2\frac{\dd^2}{\dd z^2} + z\frac{\dd}{\dd z} -\frac{1}{4}\Big] J_\alpha(z)=
c^{2}\Big[\frac{\alpha^2 -\frac{1}{4}}{z^2} - 1\Big] K_\alpha(xy) \, ,
}
where in the last step we used the form of the Bessel differential equation. This is indeed the required identity.

\section{Entropy for $d \ge 2$}

Here we elaborate on the calculation of the entropy in higher dimensions. The starting point is to
understand how the result \eqref{Sl} for a given angular momentum $\ell$ emerges. As discussed in
the main text, the radial kernel of the continuum problem is dual to a discrete kernel that describes
a subsystem $[\ell+1,\infty)$ in a hopping chain with a linear potential. Here we summarize how to
obtain the entropy in the curved-space CFT framework \cite{DSVC17}. The main idea of the method is to
interpret the spatial variation of the Fermi velocity as a nonflat metric in the underlying CFT.
For the gradient chain, a local density approximation yields the effective dispersion
and corresponding Fermi velocity
\eq{
\omega_q(x) = -\cos q +\frac{x}{c}, \qquad
v_F(x) = \left.\frac{\dd \omega_q(x)}{\dd q}\right|_{q_F(x)}=
\sqrt{1-\frac{x^2}{c^2}},
\label{omvf}
}
where $q_F(x) = \arccos(x/c)$ and $c$ is the length scale created by the gradient. The goal is then
to find isothermal complex coordinates $z$, such that the Riemannian metric of the inhomogeneous chain
$\dd s^2=\ee^{2\sigma(x)} \dd z \dd \Bar{z}$ is Weyl-equivalent to the flat metric, where the Weyl
factor $\ee^{\sigma(x)}=v_F(x)$ plays the role of the Fermi velocity. It is easy to see that the proper choice is
\eq{
z= \tilde{x}+ i t, \qquad  \tilde{x}=\int_0^x \frac{\dd y}{v_F(y)} = c \arcsin(\frac{x}{c})
} 
as it yields a local rescaling $\dd t \to v_F(x) \dd t$ of time in the curved metric. Note that this transformation
brings the Euclidean spacetime from the original strip $[-c,c]\times \mathbb{R}$ into $[-\pi c/2,\pi c/2]\times \mathbb{R}$.

The entropy is obtained via the replica trick combined with the twist-field method \cite{DSVC17}.
The R\'enyi entropy
\eq{
S_{\ell,n} = \frac{1}{1-n} \ln \, \twn{\ell}_{\mathrm{curved}}
}
is given via a path integral with a twist field $\mathcal{T}_n(\ell)$ inserted at the boundary $\ell$ of the subsystem.
To calculate the expectation value in the curved space, we perform a Weyl transformation to the isothermal
coordinates which flattens out the metric. In a subsequent step, we transform the strip with the flat metric
to the upper half plane, using the mapping $g(z)=\ee^{i(z/c+\pi/2)}$. The one-point function then transforms as
\eq{
\twn{\ell}_{\mathrm{curved}} =
\ee^{-\sigma(\ell)\Delta_n} \twn{z_0}_{\mathrm{flat}} =
\ee^{-\sigma(\ell)\Delta_n} \left|\frac{\dd g}{\dd z}\right|^{\Delta_n}_{z_0} \twn{g(z_0)}_{\mathrm{UHP}},
}
where $z_0 = \tilde x(\ell)$ is the boundary coordinate in the flat metric and the scaling dimension of
the twist field is given by $\Delta_n=\frac{1}{12}(n-1/n)$. Inserting the Fermi velocity in \eqref{omvf}
for the Weyl factor and using $\twn{w}_{\mathrm{UHP}} \propto [\Im(w)]^{-\Delta_n}$ for the expectation
value on the upper half  plane, one arrives at
\eq{
S_{\ell,n} = \frac{1}{12}\frac{n+1}{n}\ln \Big[c \Big(1-\frac{\ell^2}{c^2}\Big)\Big] + C_n,
\label{Sln}
}
where $C$ is a nonuniversal constant, which is not fixed by CFT. For a homogeneous chain,
it has a nontrivial dependence on the Fermi momentum $q_F$ \cite{JK04}. To obtain the result
for the inhomogeneous setting, one has to replace it with the local Fermi momentum, $q_F \to q_F(\ell)=\arccos(\ell/c)$,
which then yields
\eq{
C_n = \frac{1}{12}\frac{n+1}{n} \ln [\sin q_F(\ell)] + \mathcal{C}_n=
\frac{1}{12}\frac{n+1}{n} \ln \sqrt{1-\frac{\ell^2}{c^2}} + \mathcal{C}_n,
\label{Cn}
}
where $\mathcal{C}_n$ is a known constant, independent of $q_F(\ell)$ \cite{JK04}.
Putting \eqref{Sln} and \eqref{Cn} together, and taking the limit $n\to1$ with $\mathcal{C}_1\equiv\mathcal{C}$,
one arrives at \eqref{Sl} in the main text.

The total entropy \eqref{Stot} is given as a sum over the various angular momentum sectors.
The leading term violates the area law
logarithmically in arbitrary dimensions. It arises from the first term in \eqref{Sl}, with the prefactor given by the sum
\eq{
\sigma_d = \frac{1}{6} \lim_{c\to\infty} \sum_{\ell=0}^{c} \frac{M_\ell}{c^{d-1}}=
\frac{1}{3(d-1)!} \, .
\label{sigd}}
The area-law prefactor $A_d$ has two contributions
corresponding to the summation of the last two terms in \eqref{Sl}. The constant piece behaves
in exactly the same way as the logarithmic one, i.e. it contributes $6\,\sigma_d \, \mathcal{C}$ to $A_d$.
To evaluate the $\ell$-dependent sum, we introduce the scaling variable $x=\ell/c$ and consider
$\ell, c \gg 1$. Using the asymptotics of the multiplicity, $M_\ell \approx 2  \ell^{d-2}/(d-2)!$,
the contribution becomes
\eq{
\frac{1}{c^{d-1}}\sum_{\ell=0}^{c} \frac{M_\ell}{4}\ln[1-(\ell/c)^2] \approx
\frac{1}{2(d-2)!}\int_{0}^{1} \dd x \, x^{d-2} \ln (1-x^2) =
-\frac{\psi(\frac{d+1}{2}) + \gamma}{2(d-1)!} \, .
}
Adding the two pieces leads to the result \eqref{sigdAd} reported in the main text.

Finally, we compare the prefactor $\eqref{sigd}$ to the general formula which was originally conjectured
in \cite{GK06}, and later proved in \cite{LSS14}. Introducing the notation $\partial F$ and $\partial R$
for the Fermi surface and the subsystem boundary, respectively, the prefactor is given by a double integral
\eq{
\sigma_d = \frac{1}{12(2\pi)^{d-1}}
\int_{\partial F}\dd \Omega_F  \int_{\partial R} \dd \Omega_R |{\mathbf n}_F{\mathbf n}_R|,
\label{GK}}
where ${\mathbf n}_F$ and ${\mathbf n}_R$ are normal vectors on the corresponding surfaces.
Note that in our case both of these are just the surface of the unit sphere, as the radii have already
been scaled out. To evaluate the integrals, we only need the formula for the $(d-1)$-dimensional
surface area of the unit sphere
\eq{
\mathcal{S}_{d-1} = \frac{2 \pi^{d/2}}{\Gamma(\frac{d}{2})} .
\label{Sd}}
Indeed, since the integrand of \eqref{GK} contains only the relative angle between two points
on the unit sphere, ${\mathbf n}_F{\mathbf n}_R=\cos \theta$, one of the integrals gives $\mathcal{S}_{d-1}$
and the other one gives $\mathcal{S}_{d-2}$ times the integral on the relative angle
\eq{
\sigma_d = \frac{1}{12(2\pi)^{d-1}} \mathcal{S}_{d-1} \mathcal{S}_{d-2}
\int_{0}^{\pi} \dd \theta \sin^{d-2} \theta |\cos \theta| \, .
\label{sigd2}}
Note that the factor $\sin^{d-2}\theta$ in the integral comes from the expression of the surface element
$\dd \Omega$ in terms of the angular coordinates, and the corresponding integral can be
evaluated explicitly as
\eq{
\int_{0}^{\pi} \dd \theta \sin^{d-2} \theta |\cos \theta| = \frac{2}{d-1} \, .
\label{int}}
Inserting \eqref{Sd} and \eqref{int} into \eqref{sigd2}, one finds the result in \eqref{sigd}.

\end{document}